\documentclass[aps,prd]{revtex4}
\usepackage{subfigure,graphics} 
\usepackage{flafter} 
\begin{document}

\title{
Balancing the vacuum energy in heterotic $M$-theory}
\author{Nasr Ahmed}
\email{nasr.ahmed@ncl.ac.uk}
\affiliation{Astronomy Department, National Research Institute of Astronomy and Geophysics, Helwan,
Egypt}
\author{Ian G. Moss}
\email{ian.moss@ncl.ac.uk}
\affiliation{School of Mathematics and Statistics, Newcastle University, NE1 7RU, UK}

\date{\today}


\begin{abstract}
Moduli stabilisation is explored in the context of low-energy heterotic $M$-theory to show that a
small value of the cosmological constant can result from a balance between the negative potential
energy left over from stabilising the moduli and a positive Casimir energy from the higher
dimensions. Supersymmetry breaking is induced by the fermion boundary conditions on the two branes
in the theory. An explicit calculation of the Casimir energy for the gravitino reveals that the
energy has the correct sign, although the size of the contribution is close to the edge of the
parameter range for which the calculation is valid.
\end{abstract}
\pacs{PACS number(s): }

\maketitle
\section{introduction}

Horava and Witten \cite{Horava:1995qa,Horava:1996ma} proposed some time ago that the low-energy
limit of strongly coupled $E_8\times E_8$ heterotic string theory could be formulated as eleven
dimensional supergravity on a manifold with boundary. This was an important step on the road to
$M$-theory, and heterotic $M$-theory is now regarded as one of the low-energy limits of $M$-theory
and a possible link between $M$-theory and particle phenomenology \cite{Witten:1996mz,banks96}.

The original formulation of heterotic $M$-theory was based on 11-dimensional supergravity and then
built up order by order as a series in a small parameter $\epsilon\sim\kappa_{11}{}^{2/3}$
depending on the gravitational coupling $\kappa_{11}$.
At the next order came two 10-dimensional boundary terms with
$E_8$-Yang-Mills matter multiplets. Unfortunately, at higher orders the theory became ill-defined.
This problem was resolved several years later by a simple modification to
the boundary conditions, resulting in a low-energy effective theory which is supersymmetric to all
orders. We
shall be using this improved version of heterotic $M$-theory, described in Refs.
\cite{Moss:2003bk,Moss:2004ck,Moss:2005zw,Moss:2008ng}.

For applications to particle phenomenology, six of the internal dimensions lie on a deformed
Calabi-Yau
manifold and one internal dimension stretches between the two boundaries containing the matter
fields \cite{Witten:1996mz,banks96,lukas98,Lukas:1998tt}. A
variety of mechanisms have been proposed for stabilising the large number of moduli resulting from
this compactification:
\begin{itemize}
\item internal fluxes, which might fix the $(2,1)$ type of Calabi-Yau moduli by analogy with the
moduli stabilisation used in type IIB string theory \cite{Kachru:2002he,Kachru:2003aw}
\item membranes stretching between the boundaries, which contribute to fixing the $(1,1)$ type of
Calabi-Yau moduli
\cite{Curio:2001qi,Buchbinder:2003pi,Braun:2006th}
\item gaugino condensation, which contributes to fixing the Calabi-Yau volume
\cite{Dine:1985rz,Horava:1996vs,Lukas:1997rb,Becker:2004gw,Gray:2007qy}
\end{itemize}
In addition to the moduli stabilisation, an `uplifting' mechanism has to be found which makes the
effective cosmological constant non-negative \cite{Kachru:2003aw}. Fully consistent reductions of
heterotic $M$-theory have been constructed along these lines, with anti-5-branes providing an
uplifting mechanism \cite{Braun:2006th}. These reductions have also taken advantage of the extra
flexibility allowed by the inclusion of 5-branes.

In this paper we explore the possibility that the small value of
the cosmological constant results from a balance between the negative potential energy
left over from stabilising the moduli and a positive Casimir energy from the higher dimensions.
To be more specific, we use the difference in Casimir energy between a supersymmetric 
state and a broken symmetry state, because differences in vacuum energy can be calculated
more reliably using only the low energy effective theory. 

Some form of supersymmetry breaking is required by any uplifting mechanism, and the
possibility we consider is that the fermion chirality condition is different on the two boundaries. 
This type of supersymmetry breaking was first introduced into heterotic $M$-theory by Antoniadis and
Queros \cite{Antoniadis:1997ic}. They argued that modifying the fermion boundary conditions was
analagous to introducing a gaugino condensate into the weakly coupled superstring theory.  The new
formulation of heterotic $M$-theory \cite{Moss:2003bk,Moss:2004ck,Moss:2005zw,Moss:2008ng}
 makes the link between gaugino condensation and the boundary
conditions on the fermion fields explicit, so that this type of supersymmetry
breaking is realised spontaneously whenever a gaugino condensate is present.

In the past, Casimir energy contributions have been widely used to fix the values of radion modulus
field in brane-world models \cite{Toms:2000bh,Garriga:2000jb,Garriga:2001ar,Flachi:2003bb}. These
calculations have also been extended to models with modified fermion boundary conditions
\cite{Fabinger:2000jd,Flachi:2001ke,Flachi:2001bj}. In the present work, the Casimir force between
the branes is relatively small compared to the moduli stabilisation effects described above, but
the Casimir vacuum energy has a similar scale to the negative potential energy. Just as
with the type IIB superstring \cite{Kachru:2003aw}, one of the parameters in the theory has to be
fine-tuned to obtain a small cosmological constant.

The relevant features of the new formulation of heterotic $M$-theory are described in the next
section, where we also consider the backgrounds and moduli for the reduced theory. Moduli
stabilisation and the vacuum energy are covered in Sect. \ref{ms}. A calculation of the
Casimir energy for the gravitino is done in Sect. \ref{ve}.

In this paper, the 11-dimensional coordinate indices are denoted by $I,J,\dots$, and the coordinate
indices on the boundary are denoted by $A,B,\dots$ with the outward normal index by $N$. Otherwise,
conventions generally follow Green, Schwarz and Witten \cite{Green:1987mn}. In particular, the
11-dimensional gamma matrices satisfy  $\{\Gamma_I,\Gamma_J\}=2  g_{IJ}$, where the metric
signature is 
$-+\dots +$. In the reduced theory, four dimensions are indicated by $\mu,\nu\dots $ and the fifth
by $z$.

\section{Heterotic $M$-theory}
\label{hmt}

Heterotic $M$-theory is formulated on a manifold ${\cal M}$ with a boundary consisting of two
disconnected components ${\cal M}_1$ and ${\cal M}_2$. The eleven-dimensional part of
the action is conventional for supergravity,
\begin{eqnarray}
S_{SG}=&&{1\over 2 \kappa_{11}^2}\int_{\cal M}\left(-R(\Omega)
-\bar\Psi_I\Gamma^{IJK}D_J(\Omega^*)\Psi_K-\frac1{24}
G_{IJKL}G^{IJKL}\right.
\nonumber\\
&&\left.-\frac{\sqrt{2}}{96}\left(\bar\Psi_I\Gamma^{IJKLMP}\Psi_P
+12\bar\Psi^J\Gamma^{KL}\Psi^M\right)G^*_{JKLM}
-\frac{2\sqrt{2}}{11!}
\epsilon^{I_1\dots I_{11}}(C\wedge G\wedge G)_{I_1\dots I_{11}}\right)dv,
\label{actionsg}
\end{eqnarray}
where $\Psi_I$ is the gravitino, $G$ is the abelian field strength and $\Omega$ is the tetrad
connection. The combination $G^*=(G+\hat G)/2$, where hats denote the standardised subtraction of
gravitino terms to make a supercovariant expression. 

The boundary terms which make the action supersymmetric are \cite{luckock89},
\begin{equation}
S_0={1\over  \kappa_{11}^2}\int_{\cal\partial M}\left(
\hat K\mp\frac14\bar\Psi_A\Gamma^A\Gamma^B\Psi_B\right)dv,
\end{equation}
where $K$ is the extrinsic curvature of the boundary. We shall take the  upper sign on the
boundary component ${\cal M}_1$ and the lower sign on the boundary component 
${\cal M}_2$. 

For an anomaly-free theory, we have to include further boundary terms with $E_8$-Yang-Mills
multiplets
\begin{equation}
S_{YM}=-{\epsilon\over \kappa_{11}^2}\int_{\cal \partial M}
\left(\frac14{\rm tr}F^2+
\frac12{\rm tr}\bar\chi\Gamma^AD_A(\hat\Omega^{**})\chi
+\frac14\bar\Psi_A\Gamma^{BC}\Gamma^A{\rm tr}{F}^*_{BC}\chi\right)dv,
\label{action1}
\end{equation}
where $\Omega^{**}=(\Omega+\Omega^*)/2$. The original formulation of Horava
and Witten contained an extra `$\chi\chi\chi\Psi$' term, but this term is not present in the new
version. The new theory also has boundary terms depending on $R^2$ \cite{Moss:2008ng}.

The specification of the theory is completed by a supersymmetric set of boundary
conditions. For the tangential anti-symmetric tensor components \cite{Moss:2004ck,Moss:2008ng},
\begin{equation}
C_{ABC}=\mp \frac{\sqrt{2}}{12}\epsilon\,\left(\omega^Y_{ABC}-\frac12\omega^L_{ABC}\right)
\mp\frac{\sqrt{2}}{48}\epsilon\,{\rm tr}\bar\chi\Gamma_{ABC}\chi.\label{cbc}
\end{equation}
where $\omega^Y$ and $\omega^L$ are the Yang-Mills and Lorentz Chern-Simons forms respectively.
These boundary
conditions replace a modified Bianchi identity in the old formulation. A suggestion along these
lines was made in the original paper of Horava and Witten \cite{Horava:1996ma}. Boundary conditions
on the gravitino are,
\begin{equation}
\Gamma^{AB}\left(P_\pm\pm\epsilon\Gamma P_\mp \right)\Psi_A=
\pm\epsilon\left( J_Y{}^A-\frac12J_L{}^A\right),\label{gbc}
\end{equation}
where $P_\pm=(1\pm \Gamma_N)/2$ project onto chiral spinors (using the outward-going normals) and
\begin{equation}
\Gamma=\frac1{96}{\rm tr}(\bar\chi\Gamma_{ABC}\chi)\Gamma^{ABC}.\label{gamdef}
\end{equation}
$J_Y$ is the Yang-Mills supercurrent and $J_L$ is a gravitino analogue of the Yang-Mills
supercurrent. 

The resulting theory is supersymmetric {\it to all orders} in the parameter $\epsilon$ when working
to order $R^2$ in the curvature. The gauge, gravity and supergravity anomalies all vanish if
\begin{equation}
\epsilon={1\over 4\pi}\left({\kappa_{11}\over4\pi}\right)^{2/3}.\label{epsval}
\end{equation}
Further details of the anomaly cancellation, and additional Green-Schwarz terms, can be found in
Ref. \cite{Moss:2005zw}.

\subsection{Background}

Reduction of low-energy heterotic $M-$theory to  $4$-dimensions follows a traditional route, where
the light fields in the  $4-$dimensional theory correspond to the moduli of a background solution
in $11$-dimensions. The anzatz for the background metric is based on a warped product $M\times
S^1/Z_2\times Y$ where $Y$ is a Calabi-Yau space 
\cite{Witten:1996mz,lukas98,Lukas:1998tt,Lukas:1998ew}.  Since $S^1/Z_2$ is topologically the same
as a finite interval, there are two copies of the $4-$dimensional manifold $M$, the visible brane
$M_1$ and the hidden brane $M_2$, separated by a distance $l_{11}$.  A typical value for
the inverse radius of the Calabi-Yau space would be of order the Grand Unification scale
$10^{16}$GeV and the inverse separation would be of order $10^{14}$GeV.

The background solutions are only determined order by order in $\epsilon$. The improved version of
heterotic $M$-theory, which starts from an action which is
valid to all orders in $\epsilon$, gives better control of the error terms than previous versions of
the theory. We shall follow Ref. \cite{Ahmed:2008jz} for the reduction, and further details can be
found in that reference.

The explicit form of the background metric anzatz which we shall use is
\begin{equation}
ds^2=V^{-2/3}dz^2+V^{-1/3}\Phi^2\tilde g_{\mu\nu}dx^\mu dx^\nu+
V^{1/3}(\tilde g_{a\bar b}dx^adx^{\bar b}+\tilde g_{\bar ab}dx^{\bar a}dx^b),\label{rmetric}
\end{equation}
where $\tilde g_{a\bar b}$ a fixed metric on $Y$ and $V\equiv V(z,x^\mu)$, $z_1\le z\le z_2$ is
related to both the volume of the Calabi-Yau space and the warping of the 5-dimensional part of the
metric. Our background metric anzatz is similar to one used by Curio and
Krause \cite{Curio:2000dw}, except that we use a different coordinate $z$ in the $S^1/Z_2$
direction. The factor $\Phi^2$ is required to put the put the metric $\tilde g_{\mu\nu}$ on $M$ into
the Einstein frame.

Initially, we restrict the class of Calabi-Yau spaces to those with only one harmonic
$(1,1)$ form, $h_{1,1}=1$. In this case the background flux for the
antisymmetrc tensor field depends on only one parameter $\alpha$,
\begin{equation}
G^0_{ab\bar c\bar d}=\frac13\alpha\left(\tilde g_{a\bar c}\tilde g_{b\bar d}-
\tilde g_{a\bar d}\tilde g_{b\bar c}\right)\label{gflux}
\end{equation}
This anzatz solves the field equation $\nabla\cdot G^0=0$. The  boundary conditions (\ref{cbc}) are
satisfied if there is a non-zero Yang-Mills flux on the hidden brane and,
\begin{equation}
\alpha={4\sqrt{2}\pi^2\over v_{CY}^{2/3}}\epsilon\beta\label{alpha}
\end{equation}
where $v_{CY}$ is the volume of the Calabi-Yau space and $\beta$ is an integer characterising the
Pontrjagin class of the Calabi-Yau space.

The volume function $V(z)$ is determined by the exact solution of the `$zz$'
component of the Einstein equations \footnote{
Our solution for $V$ is equivalent to the one used by Lukas et al.  in ref. \cite{Lukas:1998tt} when
adapted to our coordinate system. They express the solution as $V=b_0H^3$. It is also equivalent to
the background used by Curio and Krause in ref. \cite{Curio:2000dw}, $V=(1-{\cal S}_1 x^{11})^2$,
when their  ${\cal S}_1=\alpha V_1^{-2/3}/\sqrt{2}$.
},
\begin{equation}
V(z)=1-\sqrt{2}\alpha z.\label{v}
\end{equation}
The background metric is then consistent with all of the Einstein equations apart from the ones with
components along the Calabi-Yau direction, where the Einstein tensor vanishes but the stress energy
tensor is $O(\alpha^2)$. The difference between an exact solution to the Einstein equations and the
metric anzatz $\delta g_{IJ}=O(\alpha^2)$. If we calculate the action to reduce the theory
to four dimensions, then the error in the action is $O(\alpha^4)$. As long as we work within this
level of approximation we can use the Calabi-Yau approximation as the background for our reduced
theory. Note that this approximation is uniform in $z$, and having small values of $\alpha$ does not
necessarily mean small values of $\alpha z$.

This background metric generalises for any internal Calabi-Yau space and accordingly is known as the
universal solution \cite{Lukas:1998tt}. The $h_{1,1}$ integers $\beta_i$ which characterise the
Pontrjagin class of the Calabi-Yau space can be combined into a single parameter $\beta$,
defined by
\begin{equation}
\beta=(\beta^i\beta_i)^{2/3}
\end{equation}
where $d_{ijk}\beta^i\beta^j=6\beta_i$ and $d_{ijk}$ are Calabi-Yau intersection numbers. The
parameter $\alpha$ is defined by Eq. (\ref{alpha}). Generally,  $\beta$ is not an integer, but most
examples have $\beta\ge1$.  In cases where $h_{1,1}>1$ there are other solutions with different
functional behaviour for $V(z)$ but we shall only consider the universal solution.

We shall be focussing especially on two moduli of the 4-dimensional theory, the values of $V$
on the two boundaries, $V_1$ and $V_2$. When $V_1$ and $V_2$ depend on $x^\mu$, the factor $\Phi^2$
required to put the metric 
$\tilde g_{\mu\nu}$ into the Einstein frame is,
\begin{equation}
\Phi=\left(V_1^{4/3}-V_2^{4/3}\right)^{-1/2}.\label{phi}
\end{equation}
With this definition of the Einstein metric, the gravitational coupling in 4 dimensions is given by
\begin{equation}
\kappa_4^2={4\sqrt{2}\over 3}{\kappa^2_{11}\over v_{CY}}\alpha.\label{kappa4}
\end{equation}

The reduction to 4 dimensions has also been done using a superfield formalism by Correia et
al \cite{Correia:2006pj}. This shows that in the $h_{1,1}=1$ case the reduced theory is a
supergravity model with $V_1$ and $V_2$ belonging to chiral superfields $S_1$ and $S_2$ with Kahler
potential
\begin{equation}
K=-3\ln\left((S_1+S_1^*)^{4/3}-(S_2+S_2^*)^{4/3}\right)\label{kpot}
\end{equation}
Note that, for the real scalar components, the conformal factor introduced in Eq. (\ref{rmetric})
and the Kahler potential are related by
\begin{equation}
\Phi=2^{2/3}e^{K/6}.\label{phik}
\end{equation}
In the weakly-coupled superstring limit, which corresponds to small brane-separation, $S_1\approx
S+\alpha T/\sqrt{2}$ and $S_2\approx S-\alpha T/\sqrt{2}$, where $S$ and $T$ are heterotic
superstring moduli.

\subsection{Energy scales}

The metric anzantz for the universal solution can also be used in the Yang-mills action on the
boundary $\partial{\cal M}$,
\begin{equation}
S_{YM}=-{\epsilon v_{CY}\over 2\kappa_{11}^2}\int_{\partial \cal M}\frac 14V {\rm tr}(F^2)d\tau.
\end{equation}
This suggests that the grand-unification fine-structure constant is related to a parameter 
$\alpha_{GUT}$,
\begin{equation}
\alpha_{GUT}={\kappa_{11}^2\over 8\pi\epsilon v_{CY}}.\label{gym}
\end{equation}
(The grand-unification fine-structure constant is actually $\alpha_{GUT}/V_1$). Eqs. (\ref{epsval}),
(\ref{kappa4}), (\ref{alpha}) and (\ref{gym}) allow all of the model parameters $\kappa_{11}$,
$\epsilon$, $\alpha$ and $v_{CY}$ to be expressed in terms of the topological parameter $\beta$, the
Planck length $\kappa_4$ and the fine-structure parameter $\alpha_{GUT}$:
\begin{eqnarray}
v_{CY}^{-1/6}&=&\left(2\over 3\pi\right)^{1/2}\beta^{1/2}\alpha_{GUT}\kappa_4^{-1}\label{vscale}\\
\kappa_{11}^{-2/9}&=&\left(2\over 3\pi\right)^{1/2}\left({\sqrt{2}\pi}\right)^{-1/9}
\beta^{1/2}\alpha_{GUT}^{5/6}\kappa_4^{-1}\\
\alpha&=&\frac12\left(2\over 3\pi\right)^{1/2}\beta^{3/2}\alpha_{GUT}^{3/2}\kappa_4^{-1}
\label{ascale}
\end{eqnarray}
These show clearly how choosing a self-consistent background leaves very little freedom in the
choice
of scales. For 4-dimensional Planck scale around $\kappa_4^{-1}=2.44\times 10^{18}$GeV,
$\alpha_{GUT}=0.04$ and $\beta=1$,  the Calabi-Yau energy scale becomes 
$4.5\times 10^{16}$GeV and the brane separation scale 
$l_{11}^{-1}\sim \alpha=9.0\times10^{14}$GeV.

Another scale of interest later on is the
brane separation in the $5$-dimensional Einstein metric,
\begin{equation}
l_5={1\over \sqrt{2}\alpha}(V_1-V_2).\label{l5}
\end{equation}
When compared to the radius of the Calabi-Yau space,
\begin{equation}
{v_{CY}^{1/6}\over l_5}={1\over\sqrt{2}}{\beta\alpha_{GUT}^{1/2}\over V_1-V_2}.
\end{equation}
The separation in the fifth dimension is larger than the radius of the Calabi-Yau space
when the right-hand-side of this equation is small. However, this need not be
the case when $V_1-V_2=O(\alpha_{GUT}{}^{1/2})$, which we shall call small warping.
(This is not the same as the weakly-coupled superstring limit 
$V_1-V_2\ll \alpha_{GUT}{}^{1/2}$).
Situations with small warping have to be handled carefully, with a consideration of the masses of
the Kaluza-Klein states. 

\subsection{Condensates and fluxes}

Fermion condensates and fluxes of antisymmetric tensor fields may both play a role in the
stabilisation of moduli fields. In the context of low energy heterotic $M$-theory the most likely
candidate for forming a fermion condensate is the gaugino on the hidden brane, since the effective
gauge coupling on the hidden brane is larger and runs much more rapidly into a strong coupling
regime than the gauge coupling on the visible brane.

The anzatz for a gaugino condensate on the boundary $M_i$ is \cite{Dine:1985rz},
\begin{equation}
\langle{\rm tr}\,\bar\chi_i\Gamma_{abc}\chi_i\rangle=\Lambda_i\omega_{abc}\label{cond}
\end{equation}
where $\Lambda_i$ depends only on the modulus $V_i$ and $\omega_{abc}$ is the covariantly constant
$3-$form on the Calabi-Yau space (i.e. the one with volume $v_{CY}$).
In the improved formulation of low energy heterotic $M$-theory, gaugino condensates act as sources
for a contribution $G_g$ to the field strength through the boundary conditions.
The new flux contribution is
\begin{equation}
G_{g\ abcz}=-\frac{\alpha}3(\Lambda_1+\Lambda_2)
\epsilon\Phi^2\omega_{abc}V^{1/3}.\label{gabc}
\end{equation}
This flux term give rise to a superpotential \cite{Ahmed:2008jz},
\begin{equation}
W_g=-3\sqrt{2}\alpha \epsilon\,(\Lambda_1+\Lambda_2)\label{superpot}
\end{equation}
where $\Lambda_1$ and $\Lambda_2$ are the amplitudes of the condensates (\ref{cond}).
Note that this formula works equally well for large as well as small warping.

\section{Moduli stabilisation with vanishing cosmological constant}
\label{ms}

Moduli stabilisation can be achieved by following a similar pattern to moduli stabilisation in type
IIB string theory \cite{Kachru:2003aw}. The first stage involves finding a suitable superpotential
which fixes the moduli but leads to an Anti-de Sitter vacuum. The negative energy of the vacuum
state is then raised by adding a non-supersymmetric contribution to the energy. Finally, one of the
parameters in the superpotential is fine-tuned to make the total vacuum energy very small. This
last step is justified by the plethora of Calabi-Yau metrics and fluxes which gives us a wide
range of parameters to choose from.  

The potential is given in terms of the Kahler potential $K$ and the superpotential $W$,
\begin{equation}
V=\kappa_4^{-2}e^K\left(
g^{i\bar \jmath}(D_iW)(D_j W)^*-3WW^*
\right),
\end{equation} 
where $g_{i\bar\jmath}$ is the hessian of $K$ and
\begin{equation}
D_iW=e^{-K}\partial_{V_i}(e^K\,W).
\end{equation}
Minima of the potential occur when $D_iW=0$, and the value of the potential at these minima is
always negative,
\begin{equation}
V_{\rm min}=-3\kappa_4^{-2}e^K|W|^2.\label{vgmin}
\end{equation}
If these minima exist, their location is fixed under
4-dimensional supersymmetry transformations. However, the boundary conditions at the potential
minima are not generally preserved under 5-dimensional supersymmetry and this can lead to
additional supersymmetry-breaking terms in the potential.

We shall examine the supersymmetric minima of the potential for two toy models. We shall concentrate
on general features rather than obtaining a precise fit with particle phenomenology.

\subsection{Stabilisation with condensates}

Following the type IIB route, we assume the existence of a flux term $W_f$ in the superpotential
which stabilises the $(2,1)$ moduli, and then remains largely inert whilst the other moduli are
stabilised \cite{Kachru:2002he,Kachru:2003aw}.

The gauge coupling on the hidden brane runs to large values at moderate energies and this is usually
taken to be indicative of the formation of a gaugino condensate. Local supersymmetry restricts the
form of this condensate to \cite{Burgess:1995aa}
\begin{equation}
\Lambda_2=B_2 \,v_{CY}^{-1/2}e^{-\mu V_2}\label{lam2}
\end{equation}
where $B_2$ is a constant and $\mu$ is related to the renormalisation group $\beta$-function by
\begin{equation}
\mu={6\pi\over b_0\alpha_{GUT}},\quad \beta(g)=-{b_0\over 16\pi^2}g^3+\dots.
\end{equation}
Putting in the value $b_0=90$ for the $E_8$ gauge group and the phenomenological value
$\alpha_{GUT}\approx 0.04V_1$ for the gauge coupling gives $\mu V_1\approx 5.2$.

The gauge coupling on the visible brane is supposed to run to large values only at low energies to
produce a hierarchy of energy scales, and a low energy condensate would have a negligible effect on
moduli stabilisation. There might, however,  be a separate gauge coupling from part of the $E_6$
symmetry on the visible brane which becomes large at moderate energies with a significant
condensate term. The requirement for this to happen is a large $\beta$-function, possibly arising
from charged scalar field contributions. The total superpotential for such a model would be given by
combining
$W_g$ from Eq. (\ref{superpot}) with $W_f$,
\begin{equation}
W=be^{-\mu V_2}+ce^{-\tau V_1}-w,
\end{equation}
where $w=-W_f$ and $b$, $c$ are constants, which we assume to be real but not necessarily positive.
We have control over the parameter $w$, through the choice of different Calabi-Yau manifolds and
fluxes, and some control over the values of $b$ and $c$ through the choice of $v_{CY}$ which has
so far been left arbitrary.

\begin{center}
\begin{figure}[ht]
\scalebox{1.0}{\includegraphics{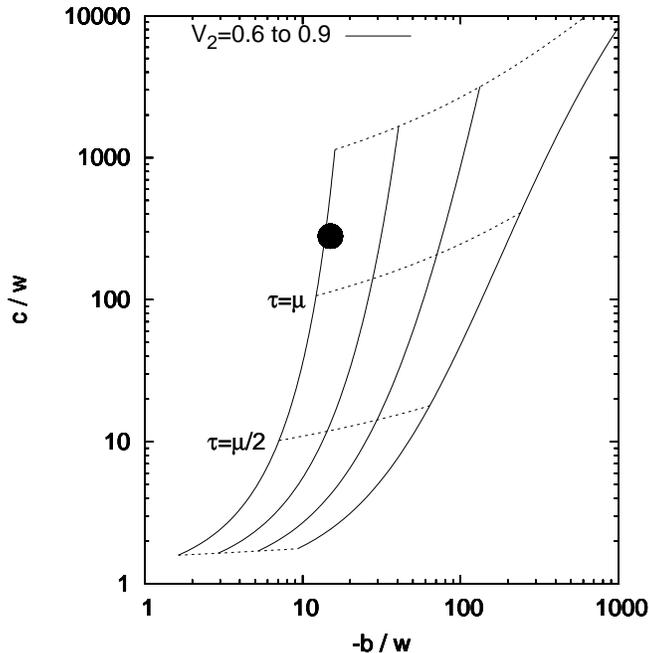}}
\caption{The contours show the value of the volume modulus $V_2$ at the supersymmetric minimum of
the potential. The axes are the superpotential parameters $b$ and $c$ and $\tau$ ranges from
$0.1\mu$ to $1.5\mu$. The spot marks the parameters used for Fig. \ref{pot2g}. This plot uses the
physical value of the volume modulus $V_1$ where $\mu V_1=5$. }
\label{contours2g}
\end{figure}
\end{center}

The moduli fields have to be compexified when evaluating the superderivatives, but for real
parameters the imaginary parts of the moduli fields play no role. The derivative terms obtained
from the Kahler
potential (\ref{kpot}) are
\begin{eqnarray}
D_1W&=&-c\tau e^{-\tau V_1}-2\Phi^2 V_1^{1/3}W,\label{d1w}\\
D_2W&=&-b\mu e^{-\mu V_2}+2\Phi^2 V_2^{1/3}W.\label{d2w}
\end{eqnarray}
At the supersymmetric minima where $D_iW=0$ we can express the parameters $b/w$ and $c/w$ in terms
of the values of $V_1$ and $V_2$ and obtain the diagram shown in Fig. \ref{contours2g}.
These are local minima, but they are the only minima which survive after adding in the extra terms
to the potential described in the next section. The potential becomes infinite when the moduli are
equal and and tends to zero when either modulus tends to infinity.

The values of the potential at the supersymmetric minimum can be expressed in terms of the
condensate scale $\Lambda=\Lambda_1+\Lambda_2$  using Eqs. (\ref{phik}), (\ref{superpot}),
(\ref{vgmin}), (\ref{d1w}) and (\ref{d2w}),
\begin{equation}
V_{\rm min}=-{27\over 32}\left({\mu\tau\Phi\over\tau V_2^{1/3}-\mu
V_1^{1/3}}\right)^2\kappa_4^{-2}\alpha^2\epsilon^2\Lambda^2.
\label{vmin}
\end{equation}
To give some idea of the magnitude of the potential, using Eq. (\ref{lam2}) for the condensate scale
and Eqs. (\ref{vscale}-\ref{ascale}) for the other parameters implies that 
$\kappa_4^{-2}\alpha^2\epsilon^2\Lambda_2^2\approx(3.6\times 10^{14}\hbox{GeV})^4B_2^2$.

\subsection{The Casimir energy contribution}

Turning on the gaugino condensate breaks the five-dimensional supersymmetry by changing the
boundary conditions and contributing to particle masses. We shall refer to these
supersymmetry breaking-boundary conditions as `twisted' boundary conditions.  The theory will
develop a non-zero
quantum contribution to the vacuum energy as a result of the 5-dimensional supersymmetry breaking.
It is important that we only need consider the difference in vacuum energy between two states--
the supersymmetric and the broken supersymmetric states--so that we can consistenly use
the low energy effective theory and we know that the quantum vacuum energy of the supersymmetric
theory vanishes. In this section we shall consider the effects of this vacuum energy when the the main contribution comes from the low mass five-dimensional  supermultiplets. We shall also restrict attention to the case where the warping is small.

An explicit calculation of the vacuum energy is given in the next section, but for the 
present some general considerations will suffice. The Casimir energy for twisted fields depends on
the size of the extra dimension $l_5$ and the amount of twisting set by the condensate scale
$\Lambda$. The energy for a flat extra dimension (i.e. a simple product metric) is  proportional to
$l_5^{-4}$, but it has to be scaled to
the 4-dimensional Einstein frame (see Eq. (\ref{rmetric})),  which gives an extra factor
$V_1{}^{2/3}\Phi^{4}$. Generally, the quantum contribution to the vacuum energy is
\begin{equation}
V_c=V_1^{2/3}\Phi^4l_5^{-4}f(\epsilon\Lambda V_1^{-1/2})
\end{equation}
where $\epsilon\Lambda$ is a dimensionless combination of the condensate scale and the theory
parameter $\epsilon$. The function $f(\theta)$ depends on the details of the particle
supermultiplets, and vanishes at $\theta=0$ where the theory is supersymmetric. For 
small $\theta$, $f(\theta)\approx C\theta^2$, where $C$ is a constant for a chosen reduction.
After making use of Eqs. (\ref{phi}) and (\ref{l5}), the quantum vacuum energy takes the form
\begin{equation}
V_c\approx \frac94CV_1^{-1}(V_1-V_2)^{-6}\alpha^4\epsilon^2\Lambda^2,\label{vcas}
\end{equation}
when the warping is small. Superficially, this appears to be $O(\alpha^4)$, but for small warping
$V_1-V_2=O(\alpha_{GUT}{}^{1/2})$ and the quantum vacuum energy is actually $O(\alpha^2)$.

\begin{center}
\begin{figure}[ht]
\scalebox{0.5}{\includegraphics{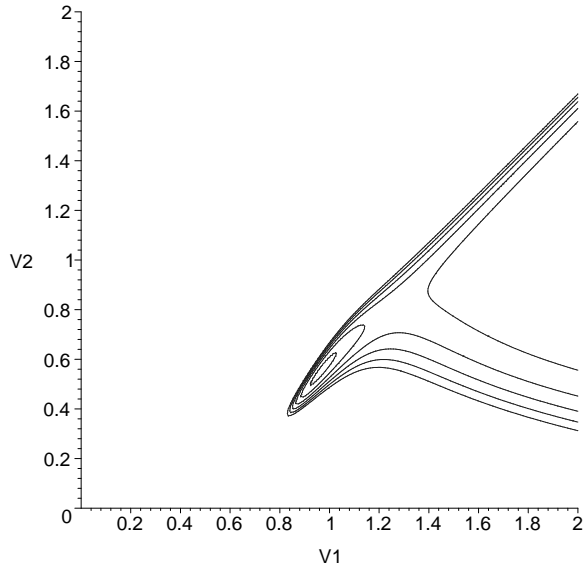}}
\caption{The contours show the value of the total potential including two gaugino condensates and
the Casimir energy with the superpotential parameters given in Fig. 1. There is a minimum with zero
potential and a saddle. The contour interval in this example is $\kappa_4^{-2}|W_f|^2$. }
\label{pot2g}
\end{figure}
\end{center}

The value of the condensate potential at its minimum is given by Eq. (\ref{vmin}), which reduces to
\begin{equation}
V_{\rm min}\approx-{81\over 128}\left({\mu\tau\over \tau-\mu}\right)^2V_1^{-1}
(V_1-V_2)^{-1}\kappa_4^{-2}\alpha^2\epsilon^2\Lambda^2,\label{vminap}
\end{equation}
when $V_1-V_2$ is small. The total vacuum energy vanishes when $V_c+V_{\rm min}=0$, which requires
choosing values of $b$, $c$ and $w$ such that
\begin{equation}
(V_1-V_2)^5={32C\over 9}\left({ \tau-\mu\over\mu\tau}\right)^2\alpha^2\kappa_4^2.\label{vcan}
\end{equation}
This is always possible as long as $\tau\ne\mu$. A more accurate numerical treatment has been used 
in Fig. \ref{pot2g} to take account of the small shift in position of the minimum when the Casimir
energy is included. The potential generally has a single minimum and a single saddle point.

The example shown in Fig. \ref{pot2g} has a very large Casimir energy contribution. For smaller
Casimir energies, Eq. (\ref{vcan}) implies that the separation in the fifth dimension is similar in
size to the Calabi-Yau radius and we have to consider carefully whether other Kaluza-Klein states
contribute to the vacuum energy. The least massive Kaluza-Klein states have a mass $m_{CY}$ related
to the first non-zero eigenvalue $\lambda_1$ of the second order perturbation operators on the
Calabi-Yau space,
\begin{equation}
m_{CY}^2=\lambda_1 v_{CY}^{-1/3}.
\end{equation} 
If the total vacuum energy vanishes, then Eq. (\ref{vcan}) holds and using Eqs.
(\ref{vscale}-\ref{ascale}) we have
\begin{equation}
l_5m_{CY}={2\sqrt{2}\over 3}\left({9C\over 2\pi}\right)^{1/5} 
\left({ \tau-\mu\over\mu\tau\beta}\right)^{2/5}\alpha_{GUT}^{1/10}
\,\lambda_1^{1/2}
\approx 0.39C^{1/5}\lambda_1^{1/2}.\label{lm}
\end{equation}
The massive Kaluza-Klein states can be neglected when $l_5m_{CY}\gg1$. This can happen if $C$ is
large or if the Calabi-Yau background  has a relatively large first eigenvalue.

The values of the first eigenvalue of the scalar Laplacian on a Calabi-Yau space have been evaluated 
for specific examples using numerical methods by Braun et. al. \cite{Braun:2008jp}. The values
range between around $20$ for a Fermat quintic to around $100$ for some more complicated cases. The
larger values are already marginally consistent with a five-dimensional reduction unless $C$
happens to be very small (or indeed negative).  It would be interesting to find out whether even
larger values can be realised, especially for large values of the topological index $\beta$.

If the Kaluza-Klein modes contribute significantly to the vacuum energy, then the Casimir energy 
might still be able raise the minimum of the potential, but a more 
sophisticated calculation of the vacuum energy is required. Another possibility is that the warping
is large, in which case a curved space casimir calculation is required.

\subsection{Stabilisation with non-perturbative terms}

If there are no high energy condensates on the visible brane, then we can replace the condensate
on the visible brane with another non-perturbative effect. The usual candidate for this is a
membrane which
stretches between the two boundaries. The area of the membrane $\propto V_1-V_2$ and the type of
contribution this gives to the superpotential is
\begin{equation}
W_{np}=ce^{-\tau(V_1-V_2)}.
\end{equation}
The total superpotential for the toy model is given by
\begin{equation}
W=be^{-\mu V_2}+ce^{-\tau(V_1-V_2)}-w,
\end{equation}
where $w=-W_f$ and $b$, $c$ are constants. 

\begin{center}
\begin{figure}[ht]
\scalebox{1.0}{\includegraphics{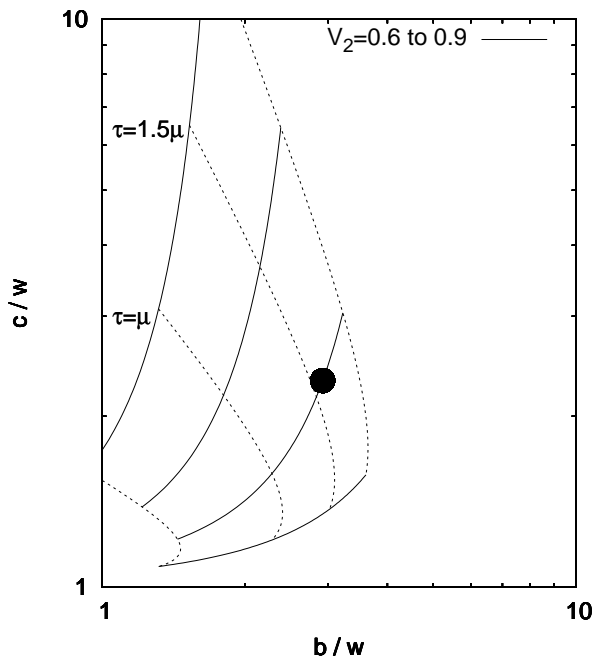}}
\caption{The contours show the value of the volume modulus $V_2$ at the supersymmetric minimum of
the potential. The axes are the superpotential parameters $b$ and $c$ and values of $\tau$ range
from $\mu/2$ to $2\mu$. In this case $c$ and $\tau$ parameterise the membrane contribution
$W_{np}$. The spot marks the parameters used for Fig. \ref{potnp}. This plot uses
the physical value of the volume modulus $V_1$ where $\mu V_1=5$.  }
\label{contours}
\end{figure}
\end{center}

This time the parameters $b$, $c$ and $w$ given in terms of the values of $V_1$ and $V_2$ at
the supersymmetric minimum are shown in figure \ref{contours}. There are always supersymmetric
minima in the parameter region indicated on the figure.
The minimum potential is now given by
\begin{equation}
V_{\rm min}=-{27\over 32}\mu^2\Phi^2(V_1^{1/3}-V_2^{1/3})^{-2}
\kappa_4^{-2}\alpha^2\epsilon^2\Lambda^2.
\label{vgmin2}
\end{equation} 

\begin{center}
\begin{figure}[ht]
\scalebox{0.5}{\includegraphics{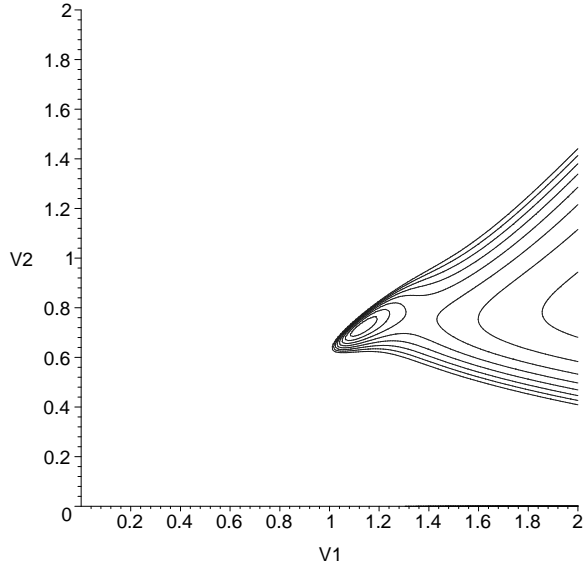}}
\caption{The contours show the value of the total potential including the supermembrane potential
and the Casimir energy using the superpotential parameters given in Fig. \ref{contours} . There is a
local minimum with zero potential and a saddle. The contour interval in this example is
$0.01\kappa_4^{-2}|W_f|^2$.}
\label{potnp}
\end{figure}
\end{center}

The same procedure as before can now be used to determine whether the Casimir energy is able to
cancel the negative gaugino induced potential. For small warping,
\begin{equation}
V_{\rm min}=-{729\over 128}\mu^2\Phi^2V_1^{-1}(V_1-V_2)^{-3}
\kappa_4^{-2}\alpha^2\epsilon^2\Lambda^2.
\label{vmin2}
\end{equation}
The five-dimensional Casimir energy (\ref{vcas}) can cancel the negative 
gaugino induced potential when
\begin{equation}
(V_1-V_2)^3={32C\over 81\mu^2}V_1^{-2}\alpha^2\kappa_4^2.
\end{equation}
This can be arranged by choosing $b$, $c$ and $w$. The potential is shown in Fig. \ref{potnp}, and
like in the previous case it has a single minimum and a single saddle point.
The size of the extra dimension satisfies
\begin{equation}
l_5m_{CY}={4\sqrt{2}\over 9}\left({9C\over 4\pi\mu^2}\right)^{1/3}\alpha_{GUT}^{1/2}
\,\lambda_1^{1/2}\approx 0.038 C^{1/3}\lambda_1^{1/2}.\label{conseq2}
\end{equation}
Either the first eigenvalue needs to be larger than in the previous case or we have to include the
Kaluza-Klein modes.

The most general situation is one in which there are combinations of different condensate-like terms
and non-perturbative terms in the superpotential. When the Calabi-Yau space has $h_{1,1}>1$ there
are additional $(1,1)$ moduli which can all be fixed at the minimum of the potential
\cite{Braun:2006th}. The general case has a potential similar to (\ref{vminap}) at the minimum, and
the balancing this against the Casimir energy gives a consistency condition which is more like Eq.
(\ref{lm}) than Eq. (\ref{conseq2}).

\section{Vacuum energy of the gravitino}
\label{ve}

After reduction from eleven to five dimensions, the theory contains a graviton multiplet and 
a scalar hypermultiplet. The hypermultiplet includes the Calabi-Yau volume field $V$. In the general
case, there is also a collection of $h_{1,1}-1$ vector multiplets and a set of $h_{2,1}$
hypermultiplets. The quantum vacuum energy of the vector multiplets was evaluated in Ref.
\cite{Moss:2004un}, and as expected there was a cancellation between the contributions from the
bosonic and fermionic fields
\footnote{There was one term left over after the cancellation which was interpreted as a
higher-order string effect. In retrospect, this term is due to the supersymmetry anomaly, and must
therefore cancel against Green-Schwarz terms.}. 
In this section, we shall consider the graviton multiplet and include the gaugino
condensate. We expect to see effects on the vacuum energy from the twisting of the gravitino
boundary conditions and from changes in the mass of the gravitino. We shall only give results in the
case where the warping of the fifth dimension is small, but even in this case we have to face some
new technical issues.

\subsection{Twisted fermion boundary conditions}

Majorana fermion fields in five dimensions have 8 components which can conveniently be
placed into an 8-spinor $\psi$. The gamma-matrix representation which we use is
\begin{equation}  
\hat\Gamma_{\mu}=\pmatrix{\gamma_\mu&0\cr 0&-\gamma_\mu\cr},\qquad
\hat\Gamma_{5}=\pmatrix{\gamma_5&0\cr 0&-\gamma_5\cr},
\end{equation}
where $\gamma_\mu$, $\mu=1\dots 4$ are the usual Dirac gamma-matrices. Following
Ref. \cite{Flachi:2001bj}, we start with identical chirality conditions on the spinor fields 
on the two flat boundary components,
\begin{equation}
\psi=\hat\Gamma_5\psi.
\end{equation}
Twisted boundary conditions are obtained by applying a similarity transformation to the
spinor representation at the hidden brane, so that the boundary condition there becomes
\begin{equation}
\psi=e^{i\theta\hat\Gamma/2}\hat\Gamma_5e^{-i\theta\hat\Gamma/2}\psi,\label{fivebc}
\end{equation}
where $\theta$ is a real number and $\hat\Gamma$ anti-commutes with the other $\gamma$-matrices, 
for example
\begin{equation}
\hat\Gamma=\pmatrix{0&1\cr1&0}.
\end{equation}
The special case $\theta=\pi$ results in a fermion with opposite chirality on the two boundaries.

In the flat space limit, the Kaluza-Klein masses for the twisted fermion modes become
$(n\pi\pm\theta/2)/l_5$, where $l_5$ is the separation of the two boundaries. The Casimir energy 
$V_F(\theta)$ for these modes is \cite{Flachi:2001ke,Flachi:2001bj}
\begin{equation}
V_F(\theta)=\pm{3\over 64\pi^2}l_5^{-4}\sum_{n=1}^{\infty}{\cos n\theta\over n^5},
\end{equation}
where the plus sign is for ordinary fermions and the minus sign for ghosts.

In the case of a supermultiplet, the total Casimir energy is the sum of Fermionic and Bosonic
contributions,
\begin{equation}
V_T=V_F(\theta)+V_B(\theta).
\end{equation}
The total Casimir energy vanishes in the supersymmetric limit where $\theta=0$. If the bosonic modes
are unaffected by the twist, then $V_B(\theta)=V_B(0)=-V_F(0)$ and the total Casimir energy in the
5-dimensional Einstein frame is given by
\begin{equation}
V_T=V_F(\theta)-V_F(0)=
\mp{3\over 32\pi^2}l_5^{-4}\sum_{n=1}^{\infty}{\sin^2(n\theta/2)\over n^5}.
\end{equation}
For small $\theta$,
\begin{equation}
V_T(\theta)\approx \mp{3\over 128\pi^2}\zeta(3) l_5^{-4}\theta^2.\label{vt}
\end{equation}
where $\zeta$ is the Riemann zeta function. The upper sign is for ordinary fermions and the lower
sign for ghosts.

\subsection{The gravitino contribution}

We shall see now that the 11-dimensional gravitino in the lightest Calabi-Yau mode is equivalent to
the twisted 5-dimensional fermion which was dealt with in the last section. The demonstration falls
into two parts. The first part follows directly from the 11-dimensional boundary conditions and the
second part of the process is to gauge away the condensate contribution to the fermion operators.
The boundary conditions and fermion operators of the 11-dimensional theory are described in the
appendix.
 
We take two flat boundary components separated in the $z=x^5$ direction labelled by the index $j=1$
for the visible brane (smaller $z$) and $j=2$ for the hidden brane (larger $z$). The 11-dimensional
spinors are all in the lightest Calabi-Yau fermion mode which is the covariantly-constant
Calabi-Yau spinor. 

The boundary conditions on the tangential gravitino components are
\begin{equation}
\left(P_\pm\pm\epsilon\Gamma P_\mp \right)\lambda_\mu=0,\label{bcstart}
\end{equation}
where $\Gamma$ was defined in Eq. (\ref{gamdef}). The upper signs are used for the visible brane,
the lower for the hidden brane, and
\begin{equation}
P_\pm=\frac12\left(1\pm\Gamma_N\right)=\frac12\left(1-\Gamma_5\right).
\end{equation}
When there is a gaugino condensate (\ref{cond}), then by considering $\Gamma^2$ it is possible to
conclude that,
\begin{equation}
\Gamma={i\over 2}\Lambda_jV_j^{-1/2}\hat\Gamma
\end{equation}
where $\hat\Gamma^2=1$ and $\{\hat\Gamma,\Gamma_\alpha\}=0$ for $\alpha=1\dots 5$. This allows us to
rewrite the boundary condition (\ref{bcstart}) on brane $j$ in the form
\begin{equation}
\lambda_\mu=e^{i\theta_j\hat\Gamma/2}\Gamma_{5}e^{-i\theta_j\hat\Gamma/2}\lambda_\mu
\end{equation}
where
\begin{equation}
\theta_j=\mp2\arctan\frac12\epsilon\Lambda_jV_j^{-1/2} \approx\mp\epsilon\Lambda_j V_j^{-1/2},
\label{thetabc}
\end{equation}
for small $\epsilon\Lambda_j$. This is equivalent to the boundary condition Eq. (\ref{fivebc}) when
written in 5-dimensional spinor form.

The gauge-ghost $\eta$ has the same boundary conditions as the tangential
gravitino, but the gauge ghost $\nu$ and the Nielsen-Kallosh ghost swaps the classes $S_+$ and $S_-$
(see the appendix). For these
\begin{equation}
c=-e^{i\theta_j\hat\Gamma/2}\Gamma_{5}e^{-i\theta_j\hat\Gamma/2}c
\end{equation}
where
\begin{equation}
\theta_j\approx\pm\epsilon\Lambda_j V_j^{-1/2}.\label{thetabc2}
\end{equation}
The normal component of the gravitino has the same value of $\theta_j$ as the tangential components.
Note that we can use a similarity transformation to reduce the value of $\theta_j$ at the visible
brane to zero, but before doing this we have to consider the fermion operators.

The fermion operators $D_m$ given in Eq. (\ref{dc}) depend on the background $G$ flux, which
consists of the brane induced part $G_0$ and the condensate induced part $G_g$ (\ref{gabc}). The
condensate part will give a contribution to the fermion determinants. Fortunately, it is possible
to gauge away the condensate terms from the operators using  
\begin{equation}
D_m=e^{-i\Theta\hat\Gamma/2}\left(\Gamma^ID_I-\frac{\sqrt{2}}{96}m\Gamma^{IJKL}G^0_{IJKL}\right)
e^{i\Theta\hat\Gamma/2}
\end{equation}
where $\Theta$ is a function of $z$,
\begin{equation}
\Theta=-\frac85m\epsilon\Lambda\Phi^2(V_1^{5/6}-V^{5/6}).
\end{equation}
This transformation transfers the effect of the gaugino condensate terms to the boundary condition
on the hidden brane, where
\begin{equation}
\Theta\approx -m\epsilon\Lambda V_1^{-1/2}\label{thetaop},
\end{equation}
for small warping. We now combine (\ref{thetabc}) or (\ref{thetabc2}) with the transformation
(\ref{thetaop}) to get
the total twist at the hidden brane
\begin{equation}
\theta=\epsilon\Lambda V_1^{-1/2}(\pm1-m).\label{tf}
\end{equation}
where the upper sign is for the gravitino (with $m=1$) and the gauge ghost $\eta$ (with $m=1/3$),
whilst the lower sign is for the gauge ghost $\nu$ (with $m=1/3$) and the Nielsen-Kallosh ghost
(with $m=-3$) . For small warping, the gravitino
components are effectively untwisted, but the two gauge-ghost fermions and the Nielsen-Kallosh ghost
survive as twisted fermions.
The total contribution to the Casimir energy (\ref{vt}) based on (\ref{tf}) is
\begin{equation}
V_T={3\over 128\pi^2}\zeta(3)\left(\frac{20}{9}+4\right) l_5^{-4}V_1^{-1}\epsilon^2\Lambda^2.
\end{equation}
In terms of the constant $C$,
\begin{equation}
V_T=Cl_5^{-4}V_1^{-1}\epsilon^2\Lambda^2.
\end{equation}
where $C\approx 1.776\times 10^{-2}$ or $C^{1/5}\approx 0.45$. The cancellations have resulted in a
small positive contribution to the vacuum energy. If the gravitino makes the only contribution to
the Casimir energy, we could satisfy the consistency condition (\ref{lm}) only when the first
eigenvalue on the Calabi-Yau space $\lambda_1\gg 32$. Alternatively, other multiplets may contribute
to the Casimir energy and reduce the consistency bound on the first eigenvalue.

\section{conclusion}

The aim of the present paper has been to show that, in principle, the Casimir energy can cancel
other contributions to the the cosmological constant in reductions of heterotic $M$-theory. The
Casimir energy arises from extra dimensions, where the fermion boundary conditions can break the
supersymmetry. Since the boundary conditions and
the moduli stabilisation potential can both be related to the scale of a gaugino condensate,
cancellation of the cosmological constant requires the cancellation of two terms with similar
scales, although the fine-tuning of a parameter in the superpotential is still required.

An explicit calculation of the Casimir energy for the gravitino revealed that the energy has the
correct sign, and the size of the contribution was marginally consistent the parameter range
for which the 5-dimensional calculation was valid. The validity of the approximation improves with
the size of the first eigenvalue of the Laplacian on the Calabi-Yau space, and so examples of
Calabi-Yau spaces with large first eigenvalues would be of interest. Two ways to improve the Casimir
calculation itself would be to allow large warping of the metric in the fifth dimension and to
extend the results to the full eleven dimensions. Some preliminary work has already been done with
large warping \cite{Ahmed}, and on the casimir energy for manifolds with topology 
$R^4\times S^1/Z_2\times S^2$ \cite{Flachi:2003bb}.

The gravitino makes the largest contribution to the casimir energy because
the gravitino boundary conditions are influenced directly by the condensate. However,
other fields may receive smaller indirect effects, through changes in the background metric 
for example, and these can also contribute towards the Casimir energy. The first step
towards calculating the contribution from these fields should be a full 5-dimensional reduction of
the new version of heterotic $M$-theory, which has not yet been done. It should be possible to
express the bulk fields of the reduced theory in terms of 5-dimensional supergravity multiplets, as
in Ref. \cite{Lukas:1998tt}, and the boundary conditions on these multiplets are currently under
investigation.

Another avenue for further work would be extending the results of this paper to include the presence
of 5-branes and anti-5-branes, which seem to be required for obtaining something approaching the
standard model of particle physics at low energies \cite{Braun:2006th}. At present, the
modifications to the boundary conditions required for 5-branes are not understood, but it seems
likely that a 5-brane running parallel to the boundary branes will lead to two independent sets of
twisted fermion boundary conditions on the boundary branes.

We turn finally to the prospects for heterotic $M$-theory cosmology. The potential for the volume
moduli has a saddle point separating the local minimum from the large volume region with zero
potential. If the potential barrier though the saddle point is sufficiently wide, an inflationary
type of evolution from the saddle to the local minimum in the potential is possible. This case also
allows an instanton  representing the `ex nihilo' creation of the universe at the saddle point
\cite{Hawking:1981fz}.    
By complete contrast, since the potential becomes positive and infinite when the brane separation
shrinks to zero, it seems to also allow a colliding brane type of cosmological scenario
\cite{Khoury:2001wf}. For this scenario, the vacuum energy would be tuned to be small at the saddle
point, rather than at the minimum of the potential. 

\appendix
\section{The gravitino in 11-dimensions}

In this appendix we shall give some details about gauge-fixing for the gravitino in eleven
dimensions, and obtain the boundary conditions on the gravitino and ghost fields.

The 2-fermion gravitino Lagrangian ${\cal L}_{RS}$ is part of the supergravity action
(\ref{actionsg}),
\begin{equation}
{\cal L}_{RS}=-\bar\Psi_I\Gamma^{IJK}D_J\Psi_K-
\frac{\sqrt{2}}{96}\left(\bar\Psi_I\Gamma^{IJKLMP}\Psi_P
+12\bar\Psi^J\Gamma^{KL}\Psi^M\right)G_{JKLM}.
\end{equation}
The action is invariant under the supersymmetry transformation
\begin{equation}
\delta\Psi_I=D_I\eta+
{\sqrt{2}\over 288}\left(\Gamma_I\Gamma^{JKLM}-12\delta_I{}^J\Gamma^{KLM}\right)G_{JKLM}\eta.
\label{dpsi}
\end{equation}
We shall replace the supersymmetry invariance with a BRST invariance by introducing gauge fixing
function
\begin{equation}
\Psi=\Gamma^I\Psi_I,\label{gff}
\end{equation}
and ghost fields. The general procedure is similar to the case of supergravity in four dimensions,
and leads to three ghosts, a pair of gauge-fixing ghosts and the Nielsen-Kallosh ghost 
\cite{Nielsen:1978mp,Kallosh:1978de}.
The 11-dimensional description below follows Freed and Moore \cite{Freed:2004yc} in general
philosophy. Note that it is not known how to close the supersymmetry algebra off-shell in eleven
dimensions, and so the BRST approach is not fully justified. We proceed under the assumption that
the one-loop results are reliable nevertheless.

It is convenient to redefine the gravitino field first by introducing
\begin{equation}
\lambda_I=\Psi_I-\frac12\Gamma_I\Psi.\label{ldef}
\end{equation}
The gravitino Lagrangian becomes
\footnote{A similar calculation was done in Ref. \cite{Lukic:2007aj}, but the final terms in
Eqs. (\ref{lrs}) and in (\ref{lrs}) are different. We can supply further details on request.}
\begin{equation}
{\cal L}_{RS}=-\bar\lambda^I D_{1}\lambda_I+\frac94\bar\Psi D_{-3}\Psi
-{\sqrt{2}\over 4}\bar\lambda^I\Gamma^{JK}\lambda^LG_{IJKL},\label{lrs}
\end{equation}
where
\begin{equation}
D_m=\Gamma^ID_I-{\sqrt{2}\over 96}m\Gamma^{IJKL}G_{IJKL}\label{dc}
\end{equation}
The convenient choice of gauge-fixing term ${\cal L}_{GF}$ is the one which simplifies the total
Lagrangian,
\begin{equation}
{\cal L}_{GF}=-\frac94\bar\Psi D_{-3}\Psi.\label{lgf}
\end{equation} 
With this choice, the ghost action ${\cal L}_{GH}$ becomes
\begin{equation}
{\cal L}_{GH}=-\bar\nu \delta\Psi-\bar c D_{-3} c,
\end{equation}
where $\nu$ and $\mu$ are gauge ghosts and $c$ is the Nielsen-Kallosh ghost. BRST invariance
requires
\begin{equation}
\delta\nu=-\frac92 D_{-3}\Psi\label{dnu}
\end{equation}
Using Eqs. (\ref{gff}) and (\ref{dpsi}) gives $\delta\Psi=D_{1/3}\eta$, and
\begin{equation}
{\cal L}_{GH}=-\bar\nu D_{1/3}\eta-\bar c D_{-3} c.
\end{equation}
The total action with Lagrangian ${\cal L}_{TOT}={\cal L}_{RS}+{\cal L}_{GF}+{\cal L}_{GH}$ is then
invariant under the BRST transformations given by Eqs. (\ref{dpsi}) and (\ref{dnu}), with all other
BRST variations vanishing.

The boundary conditions on the tangential gravitino components and the supersymmetry parameter,
given by Eq. (\ref{gbc}), are fixed by the supersymmetry, but nothing has been determined so far
about the boundary conditions on the normal component of the gravitino. Now we shall show that the
boundary conditions on the ghost fields and the normal gravitino component $\lambda_N$ are uniquely
determined by requiring them to be BRST invariant. (For an explanation of the relationship between
BRST symmetry and boundary conditions, see Ref. \cite{Moss:1996ip}).  

Consider one boundary component ${\cal M}_j$. Two sets of spinors $S_\pm$ can be defined by the
action of the projection operators of Heterotic $M$-theory,
\begin{eqnarray}
\psi\in S_-&\hbox{ if }&(P_--\epsilon\Gamma P_+)\psi=0\hbox{ on }{\cal M}_j,\\
\psi\in S_+&\hbox{ if }&(P_++\epsilon\Gamma P_-)\psi=0\hbox{ on }{\cal M}_j.
\end{eqnarray}
According to (\ref{gbc}), if there are no background supercurrents, the theory has
\begin{equation}
\Psi_A,\ \eta\in S_-.\label{evc}
\end{equation}
on the hidden brane. The operator $D_m$ maps from gauge ghosts in $S_-$ to gauge ghosts in $S_+$.
Under BRST transformations
$\delta\Psi=D_{1/3}\eta\in S_+$, therefore we conclude that $\Psi\in S_+$. The relationship between
the ghost $c$ and $\Psi$ then implies
\begin{equation}
\Psi,\ c\in S_+.\label{pvc}
\end{equation} 
The operator $D_m$ is self-adjoint on Majorana fermions, so that the BRST transformation (\ref{dnu})
implies,
\begin{equation}
\nu\in S_+
\end{equation}
For the boundary conditions on $\lambda_I$ we shall consider the case where $\Gamma$ anticommutes
with $\Gamma_I$ (which is relevant for the gaugino condensate when $I=1\dots 5$), and then from
(\ref{evc}) and (\ref{pvc}),
\begin{equation}
\lambda_A\in S_-
\end{equation}
The boundary conditions on $\lambda_N$ follow from Eq. (\ref{pvc}),
\begin{equation}
(P_++\epsilon\Gamma P_-)(\Gamma^A\Psi_A+\Gamma^N\Psi_N)=0.
\end{equation}
This implies that
\begin{equation}
(P_+-\epsilon\Gamma P_-)\lambda_N=0
\end{equation}
and we call this set of spinors $\bar S_-$. This completes a consistent set of boundary conditions,
but it would be interesting to extend these further to include the Yang-Mills fields and to
understand their mathematical significance better.

\bibliography{paper.bib}

\begin{thebibliography}{45}
\expandafter\ifx\csname natexlab\endcsname\relax\def\natexlab#1{#1}\fi
\expandafter\ifx\csname bibnamefont\endcsname\relax
  \def\bibnamefont#1{#1}\fi
\expandafter\ifx\csname bibfnamefont\endcsname\relax
  \def\bibfnamefont#1{#1}\fi
\expandafter\ifx\csname citenamefont\endcsname\relax
  \def\citenamefont#1{#1}\fi
\expandafter\ifx\csname url\endcsname\relax
  \def\url#1{\texttt{#1}}\fi
\expandafter\ifx\csname urlprefix\endcsname\relax\def\urlprefix{URL }\fi
\providecommand{\bibinfo}[2]{#2}
\providecommand{\eprint}[2][]{\url{#2}}

\bibitem[{\citenamefont{Horava and Witten}(1996{\natexlab{a}})}]{Horava:1995qa}
\bibinfo{author}{\bibfnamefont{P.}~\bibnamefont{Horava}} \bibnamefont{and}
  \bibinfo{author}{\bibfnamefont{E.}~\bibnamefont{Witten}},
  \bibinfo{journal}{Nucl. Phys.} \textbf{\bibinfo{volume}{B460}},
  \bibinfo{pages}{506} (\bibinfo{year}{1996}{\natexlab{a}}),
  \eprint{hep-th/9510209}.

\bibitem[{\citenamefont{Horava and Witten}(1996{\natexlab{b}})}]{Horava:1996ma}
\bibinfo{author}{\bibfnamefont{P.}~\bibnamefont{Horava}} \bibnamefont{and}
  \bibinfo{author}{\bibfnamefont{E.}~\bibnamefont{Witten}},
  \bibinfo{journal}{Nucl. Phys.} \textbf{\bibinfo{volume}{B475}},
  \bibinfo{pages}{94} (\bibinfo{year}{1996}{\natexlab{b}}),
  \eprint{hep-th/9603142}.

\bibitem[{\citenamefont{Witten}(1996)}]{Witten:1996mz}
\bibinfo{author}{\bibfnamefont{E.}~\bibnamefont{Witten}},
  \bibinfo{journal}{Nucl. Phys.} \textbf{\bibinfo{volume}{B471}},
  \bibinfo{pages}{135} (\bibinfo{year}{1996}), \eprint{hep-th/9602070}.

\bibitem[{\citenamefont{Banks and Dine}(1996)}]{banks96}
\bibinfo{author}{\bibfnamefont{T.}~\bibnamefont{Banks}} \bibnamefont{and}
  \bibinfo{author}{\bibfnamefont{M.}~\bibnamefont{Dine}},
  \bibinfo{journal}{Nucl. Phys.} \textbf{\bibinfo{volume}{B479}},
  \bibinfo{pages}{173} (\bibinfo{year}{1996}), \eprint{hep-th/9605136}.

\bibitem[{\citenamefont{Moss}(2003)}]{Moss:2003bk}
\bibinfo{author}{\bibfnamefont{I.~G.} \bibnamefont{Moss}},
  \bibinfo{journal}{Phys. Lett.} \textbf{\bibinfo{volume}{B577}},
  \bibinfo{pages}{71} (\bibinfo{year}{2003}), \eprint{hep-th/0308159}.

\bibitem[{\citenamefont{Moss}(2005)}]{Moss:2004ck}
\bibinfo{author}{\bibfnamefont{I.~G.} \bibnamefont{Moss}},
  \bibinfo{journal}{Nucl. Phys.} \textbf{\bibinfo{volume}{B729}},
  \bibinfo{pages}{179} (\bibinfo{year}{2005}), \eprint{hep-th/0403106}.

\bibitem[{\citenamefont{Moss}(2006)}]{Moss:2005zw}
\bibinfo{author}{\bibfnamefont{I.~G.} \bibnamefont{Moss}},
  \bibinfo{journal}{Phys. Lett.} \textbf{\bibinfo{volume}{B637}},
  \bibinfo{pages}{93} (\bibinfo{year}{2006}), \eprint{hep-th/0508227}.

\bibitem[{\citenamefont{Moss}(2008)}]{Moss:2008ng}
\bibinfo{author}{\bibfnamefont{I.~G.} \bibnamefont{Moss}},
  \bibinfo{journal}{JHEP} \textbf{\bibinfo{volume}{11}}, \bibinfo{pages}{067}
  (\bibinfo{year}{2008}), \eprint{0810.1662}.

\bibitem[{\citenamefont{Lukas et~al.}(1999{\natexlab{a}})\citenamefont{Lukas,
  Ovrut, Stelle, and Waldram}}]{lukas98}
\bibinfo{author}{\bibfnamefont{A.}~\bibnamefont{Lukas}},
  \bibinfo{author}{\bibfnamefont{B.~A.} \bibnamefont{Ovrut}},
  \bibinfo{author}{\bibfnamefont{K.~S.} \bibnamefont{Stelle}},
  \bibnamefont{and} \bibinfo{author}{\bibfnamefont{D.}~\bibnamefont{Waldram}},
  \bibinfo{journal}{Phys Rev D} \textbf{\bibinfo{volume}{59}},
  \bibinfo{pages}{086001} (\bibinfo{year}{1999}{\natexlab{a}}).

\bibitem[{\citenamefont{Lukas et~al.}(1999{\natexlab{b}})\citenamefont{Lukas,
  Ovrut, Stelle, and Waldram}}]{Lukas:1998tt}
\bibinfo{author}{\bibfnamefont{A.}~\bibnamefont{Lukas}},
  \bibinfo{author}{\bibfnamefont{B.~A.} \bibnamefont{Ovrut}},
  \bibinfo{author}{\bibfnamefont{K.~S.} \bibnamefont{Stelle}},
  \bibnamefont{and} \bibinfo{author}{\bibfnamefont{D.}~\bibnamefont{Waldram}},
  \bibinfo{journal}{Nucl. Phys.} \textbf{\bibinfo{volume}{B552}},
  \bibinfo{pages}{246} (\bibinfo{year}{1999}{\natexlab{b}}),
  \eprint{hep-th/9806051}.

\bibitem[{\citenamefont{Kachru et~al.}(2003{\natexlab{a}})\citenamefont{Kachru,
  Schulz, and Trivedi}}]{Kachru:2002he}
\bibinfo{author}{\bibfnamefont{S.}~\bibnamefont{Kachru}},
  \bibinfo{author}{\bibfnamefont{M.~B.} \bibnamefont{Schulz}},
  \bibnamefont{and} \bibinfo{author}{\bibfnamefont{S.}~\bibnamefont{Trivedi}},
  \bibinfo{journal}{JHEP} \textbf{\bibinfo{volume}{10}}, \bibinfo{pages}{007}
  (\bibinfo{year}{2003}{\natexlab{a}}), \eprint{hep-th/0201028}.

\bibitem[{\citenamefont{Kachru et~al.}(2003{\natexlab{b}})\citenamefont{Kachru,
  Kallosh, Linde, and Trivedi}}]{Kachru:2003aw}
\bibinfo{author}{\bibfnamefont{S.}~\bibnamefont{Kachru}},
  \bibinfo{author}{\bibfnamefont{R.}~\bibnamefont{Kallosh}},
  \bibinfo{author}{\bibfnamefont{A.}~\bibnamefont{Linde}}, \bibnamefont{and}
  \bibinfo{author}{\bibfnamefont{S.~P.} \bibnamefont{Trivedi}},
  \bibinfo{journal}{Phys. Rev.} \textbf{\bibinfo{volume}{D68}},
  \bibinfo{pages}{046005} (\bibinfo{year}{2003}{\natexlab{b}}),
  \eprint{hep-th/0301240}.

\bibitem[{\citenamefont{Buchbinder and Ovrut}(2004)}]{Buchbinder:2003pi}
\bibinfo{author}{\bibfnamefont{E.~I.} \bibnamefont{Buchbinder}}
  \bibnamefont{and} \bibinfo{author}{\bibfnamefont{B.~A.} \bibnamefont{Ovrut}},
  \bibinfo{journal}{Phys. Rev.} \textbf{\bibinfo{volume}{D69}},
  \bibinfo{pages}{086010} (\bibinfo{year}{2004}), \eprint{hep-th/0310112}.

\bibitem[{\citenamefont{Braun and Ovrut}(2006)}]{Braun:2006th}
\bibinfo{author}{\bibfnamefont{V.}~\bibnamefont{Braun}} \bibnamefont{and}
  \bibinfo{author}{\bibfnamefont{B.~A.} \bibnamefont{Ovrut}},
  \bibinfo{journal}{JHEP} \textbf{\bibinfo{volume}{07}}, \bibinfo{pages}{035}
  (\bibinfo{year}{2006}), \eprint{hep-th/0603088}.

\bibitem[{\citenamefont{Curio and Krause}(2002)}]{Curio:2001qi}
\bibinfo{author}{\bibfnamefont{G.}~\bibnamefont{Curio}} \bibnamefont{and}
  \bibinfo{author}{\bibfnamefont{A.}~\bibnamefont{Krause}},
  \bibinfo{journal}{Nucl. Phys.} \textbf{\bibinfo{volume}{B643}},
  \bibinfo{pages}{131} (\bibinfo{year}{2002}), \eprint{hep-th/0108220}.

\bibitem[{\citenamefont{Dine et~al.}(1985)\citenamefont{Dine, Rohm, Seiberg,
  and Witten}}]{Dine:1985rz}
\bibinfo{author}{\bibfnamefont{M.}~\bibnamefont{Dine}},
  \bibinfo{author}{\bibfnamefont{R.}~\bibnamefont{Rohm}},
  \bibinfo{author}{\bibfnamefont{N.}~\bibnamefont{Seiberg}}, \bibnamefont{and}
  \bibinfo{author}{\bibfnamefont{E.}~\bibnamefont{Witten}},
  \bibinfo{journal}{Phys. Lett.} \textbf{\bibinfo{volume}{B156}},
  \bibinfo{pages}{55} (\bibinfo{year}{1985}).

\bibitem[{\citenamefont{Horava}(1996)}]{Horava:1996vs}
\bibinfo{author}{\bibfnamefont{P.}~\bibnamefont{Horava}},
  \bibinfo{journal}{Phys. Rev.} \textbf{\bibinfo{volume}{D54}},
  \bibinfo{pages}{7561} (\bibinfo{year}{1996}), \eprint{hep-th/9608019}.

\bibitem[{\citenamefont{Lukas et~al.}(1998)\citenamefont{Lukas, Ovrut, and
  Waldram}}]{Lukas:1997rb}
\bibinfo{author}{\bibfnamefont{A.}~\bibnamefont{Lukas}},
  \bibinfo{author}{\bibfnamefont{B.~A.} \bibnamefont{Ovrut}}, \bibnamefont{and}
  \bibinfo{author}{\bibfnamefont{D.}~\bibnamefont{Waldram}},
  \bibinfo{journal}{Phys. Rev.} \textbf{\bibinfo{volume}{D57}},
  \bibinfo{pages}{7529} (\bibinfo{year}{1998}), \eprint{hep-th/9711197}.

\bibitem[{\citenamefont{Gray et~al.}(2007)\citenamefont{Gray, Lukas, and
  Ovrut}}]{Gray:2007qy}
\bibinfo{author}{\bibfnamefont{J.}~\bibnamefont{Gray}},
  \bibinfo{author}{\bibfnamefont{A.}~\bibnamefont{Lukas}}, \bibnamefont{and}
  \bibinfo{author}{\bibfnamefont{B.}~\bibnamefont{Ovrut}},
  \bibinfo{journal}{Phys. Rev.} \textbf{\bibinfo{volume}{D76}},
  \bibinfo{pages}{126012} (\bibinfo{year}{2007}), \eprint{0709.2914}.

\bibitem[{\citenamefont{Becker et~al.}(2004)\citenamefont{Becker, Curio, and
  Krause}}]{Becker:2004gw}
\bibinfo{author}{\bibfnamefont{M.}~\bibnamefont{Becker}},
  \bibinfo{author}{\bibfnamefont{G.}~\bibnamefont{Curio}}, \bibnamefont{and}
  \bibinfo{author}{\bibfnamefont{A.}~\bibnamefont{Krause}},
  \bibinfo{journal}{Nucl. Phys.} \textbf{\bibinfo{volume}{B693}},
  \bibinfo{pages}{223} (\bibinfo{year}{2004}), \eprint{hep-th/0403027}.

\bibitem[{\citenamefont{Antoniadis and Quiros}(1997)}]{Antoniadis:1997ic}
\bibinfo{author}{\bibfnamefont{I.}~\bibnamefont{Antoniadis}} \bibnamefont{and}
  \bibinfo{author}{\bibfnamefont{M.}~\bibnamefont{Quiros}},
  \bibinfo{journal}{Nucl. Phys.} \textbf{\bibinfo{volume}{B505}},
  \bibinfo{pages}{109} (\bibinfo{year}{1997}), \eprint{hep-th/9705037}.

\bibitem[{\citenamefont{Toms}(2000)}]{Toms:2000bh}
\bibinfo{author}{\bibfnamefont{D.~J.} \bibnamefont{Toms}},
  \bibinfo{journal}{Phys. Lett.} \textbf{\bibinfo{volume}{B484}},
  \bibinfo{pages}{149} (\bibinfo{year}{2000}).

\bibitem[{\citenamefont{Garriga et~al.}(2001)\citenamefont{Garriga, Pujolas,
  and Tanaka}}]{Garriga:2000jb}
\bibinfo{author}{\bibfnamefont{J.}~\bibnamefont{Garriga}},
  \bibinfo{author}{\bibfnamefont{O.}~\bibnamefont{Pujolas}}, \bibnamefont{and}
  \bibinfo{author}{\bibfnamefont{T.}~\bibnamefont{Tanaka}},
  \bibinfo{journal}{Nucl. Phys.} \textbf{\bibinfo{volume}{B605}},
  \bibinfo{pages}{192} (\bibinfo{year}{2001}), \eprint{hep-th/0004109}.

\bibitem[{\citenamefont{Garriga et~al.}(2003)\citenamefont{Garriga, Pujolas,
  and Tanaka}}]{Garriga:2001ar}
\bibinfo{author}{\bibfnamefont{J.}~\bibnamefont{Garriga}},
  \bibinfo{author}{\bibfnamefont{O.}~\bibnamefont{Pujolas}}, \bibnamefont{and}
  \bibinfo{author}{\bibfnamefont{T.}~\bibnamefont{Tanaka}},
  \bibinfo{journal}{Nucl. Phys.} \textbf{\bibinfo{volume}{B655}},
  \bibinfo{pages}{127} (\bibinfo{year}{2003}), \eprint{hep-th/0111277}.

\bibitem[{\citenamefont{Flachi et~al.}(2003)\citenamefont{Flachi, Garriga,
  Pujolas, and Tanaka}}]{Flachi:2003bb}
\bibinfo{author}{\bibfnamefont{A.}~\bibnamefont{Flachi}},
  \bibinfo{author}{\bibfnamefont{J.}~\bibnamefont{Garriga}},
  \bibinfo{author}{\bibfnamefont{O.}~\bibnamefont{Pujolas}}, \bibnamefont{and}
  \bibinfo{author}{\bibfnamefont{T.}~\bibnamefont{Tanaka}},
  \bibinfo{journal}{JHEP} \textbf{\bibinfo{volume}{08}}, \bibinfo{pages}{053}
  (\bibinfo{year}{2003}), \eprint{hep-th/0302017}.

\bibitem[{\citenamefont{Fabinger and Horava}(2000)}]{Fabinger:2000jd}
\bibinfo{author}{\bibfnamefont{M.}~\bibnamefont{Fabinger}} \bibnamefont{and}
  \bibinfo{author}{\bibfnamefont{P.}~\bibnamefont{Horava}},
  \bibinfo{journal}{Nucl. Phys.} \textbf{\bibinfo{volume}{B580}},
  \bibinfo{pages}{243} (\bibinfo{year}{2000}), \eprint{hep-th/0002073}.

\bibitem[{\citenamefont{Flachi et~al.}(2001{\natexlab{a}})\citenamefont{Flachi,
  Moss, and Toms}}]{Flachi:2001ke}
\bibinfo{author}{\bibfnamefont{A.}~\bibnamefont{Flachi}},
  \bibinfo{author}{\bibfnamefont{I.~G.} \bibnamefont{Moss}}, \bibnamefont{and}
  \bibinfo{author}{\bibfnamefont{D.~J.} \bibnamefont{Toms}},
  \bibinfo{journal}{Phys. Lett.} \textbf{\bibinfo{volume}{B518}},
  \bibinfo{pages}{153} (\bibinfo{year}{2001}{\natexlab{a}}),
  \eprint{hep-th/0103138}.

\bibitem[{\citenamefont{Flachi et~al.}(2001{\natexlab{b}})\citenamefont{Flachi,
  Moss, and Toms}}]{Flachi:2001bj}
\bibinfo{author}{\bibfnamefont{A.}~\bibnamefont{Flachi}},
  \bibinfo{author}{\bibfnamefont{I.~G.} \bibnamefont{Moss}}, \bibnamefont{and}
  \bibinfo{author}{\bibfnamefont{D.~J.} \bibnamefont{Toms}},
  \bibinfo{journal}{Phys. Rev.} \textbf{\bibinfo{volume}{D64}},
  \bibinfo{pages}{105029} (\bibinfo{year}{2001}{\natexlab{b}}),
  \eprint{hep-th/0106076}.

\bibitem[{\citenamefont{Green et~al.}(1987)\citenamefont{Green, Schwarz, and
  Witten}}]{Green:1987mn}
\bibinfo{author}{\bibfnamefont{M.~B.} \bibnamefont{Green}},
  \bibinfo{author}{\bibfnamefont{J.~H.} \bibnamefont{Schwarz}},
  \bibnamefont{and} \bibinfo{author}{\bibfnamefont{E.}~\bibnamefont{Witten}},
  \emph{\bibinfo{title}{{Superstring Theory. vol. 2: Loop amplitudes, anomalies
  and phenomenology}}} (\bibinfo{year}{1987}), \bibinfo{note}{{Cambridge
  University press, UK}. (Cambridge Monographs On Mathematical Physics)}.

\bibitem[{\citenamefont{Luckock and Moss}(1989)}]{luckock89}
\bibinfo{author}{\bibfnamefont{H.~C.} \bibnamefont{Luckock}} \bibnamefont{and}
  \bibinfo{author}{\bibfnamefont{I.~G.} \bibnamefont{Moss}},
  \bibinfo{journal}{Class Quantum Grav} \textbf{\bibinfo{volume}{6}},
  \bibinfo{pages}{1993} (\bibinfo{year}{1989}).

\bibitem[{\citenamefont{Lukas et~al.}(1999{\natexlab{c}})\citenamefont{Lukas,
  Ovrut, and Waldram}}]{Lukas:1998ew}
\bibinfo{author}{\bibfnamefont{A.}~\bibnamefont{Lukas}},
  \bibinfo{author}{\bibfnamefont{B.~A.} \bibnamefont{Ovrut}}, \bibnamefont{and}
  \bibinfo{author}{\bibfnamefont{D.}~\bibnamefont{Waldram}},
  \bibinfo{journal}{Nucl. Phys.} \textbf{\bibinfo{volume}{B540}},
  \bibinfo{pages}{230} (\bibinfo{year}{1999}{\natexlab{c}}),
  \eprint{hep-th/9801087}.

\bibitem[{\citenamefont{Ahmed and Moss}(2008)}]{Ahmed:2008jz}
\bibinfo{author}{\bibfnamefont{N.}~\bibnamefont{Ahmed}} \bibnamefont{and}
  \bibinfo{author}{\bibfnamefont{I.~G.} \bibnamefont{Moss}},
  \bibinfo{journal}{JHEP} \textbf{\bibinfo{volume}{12}}, \bibinfo{pages}{108}
  (\bibinfo{year}{2008}), \eprint{0809.2244}.

\bibitem[{\citenamefont{Curio and Krause}(2001)}]{Curio:2000dw}
\bibinfo{author}{\bibfnamefont{G.}~\bibnamefont{Curio}} \bibnamefont{and}
  \bibinfo{author}{\bibfnamefont{A.}~\bibnamefont{Krause}},
  \bibinfo{journal}{Nucl. Phys.} \textbf{\bibinfo{volume}{B602}},
  \bibinfo{pages}{172} (\bibinfo{year}{2001}), \eprint{hep-th/0012152}.

\bibitem[{\citenamefont{Paccetti~Correia
  et~al.}(2006)\citenamefont{Paccetti~Correia, Schmidt, and
  Tavartkiladze}}]{Correia:2006pj}
\bibinfo{author}{\bibfnamefont{F.}~\bibnamefont{Paccetti~Correia}},
  \bibinfo{author}{\bibfnamefont{M.~G.} \bibnamefont{Schmidt}},
  \bibnamefont{and}
  \bibinfo{author}{\bibfnamefont{Z.}~\bibnamefont{Tavartkiladze}},
  \bibinfo{journal}{Nucl. Phys.} \textbf{\bibinfo{volume}{B751}},
  \bibinfo{pages}{222} (\bibinfo{year}{2006}), \eprint{hep-th/0602173}.

\bibitem[{\citenamefont{Burgess et~al.}(1996)\citenamefont{Burgess,
  Derendinger, Quevedo, and Quiros}}]{Burgess:1995aa}
\bibinfo{author}{\bibfnamefont{C.~P.} \bibnamefont{Burgess}},
  \bibinfo{author}{\bibfnamefont{J.~P.} \bibnamefont{Derendinger}},
  \bibinfo{author}{\bibfnamefont{F.}~\bibnamefont{Quevedo}}, \bibnamefont{and}
  \bibinfo{author}{\bibfnamefont{M.}~\bibnamefont{Quiros}},
  \bibinfo{journal}{Annals Phys.} \textbf{\bibinfo{volume}{250}},
  \bibinfo{pages}{193} (\bibinfo{year}{1996}), \eprint{hep-th/9505171}.

\bibitem[{\citenamefont{Braun et~al.}(2008)\citenamefont{Braun, Brelidze,
  Douglas, and Ovrut}}]{Braun:2008jp}
\bibinfo{author}{\bibfnamefont{V.}~\bibnamefont{Braun}},
  \bibinfo{author}{\bibfnamefont{T.}~\bibnamefont{Brelidze}},
  \bibinfo{author}{\bibfnamefont{M.~R.} \bibnamefont{Douglas}},
  \bibnamefont{and} \bibinfo{author}{\bibfnamefont{B.~A.} \bibnamefont{Ovrut}},
  \bibinfo{journal}{JHEP} \textbf{\bibinfo{volume}{07}}, \bibinfo{pages}{120}
  (\bibinfo{year}{2008}), \eprint{0805.3689}.

\bibitem[{\citenamefont{Moss and Norman}(2004)}]{Moss:2004un}
\bibinfo{author}{\bibfnamefont{I.~G.} \bibnamefont{Moss}} \bibnamefont{and}
  \bibinfo{author}{\bibfnamefont{J.~P.} \bibnamefont{Norman}},
  \bibinfo{journal}{JHEP} \textbf{\bibinfo{volume}{09}}, \bibinfo{pages}{020}
  (\bibinfo{year}{2004}), \eprint{hep-th/0401181}.

\bibitem[{\citenamefont{Ahmed}(2008)}]{Ahmed}
\bibinfo{author}{\bibfnamefont{N.}~\bibnamefont{Ahmed}},
  \emph{\bibinfo{title}{{Brane worlds and low-energy heterotic $M$-theory}}}
  (\bibinfo{year}{2008}), \bibinfo{note}{{PhD} thesis}.

\bibitem[{\citenamefont{Hawking and Moss}(1982)}]{Hawking:1981fz}
\bibinfo{author}{\bibfnamefont{S.~W.} \bibnamefont{Hawking}} \bibnamefont{and}
  \bibinfo{author}{\bibfnamefont{I.~G.} \bibnamefont{Moss}},
  \bibinfo{journal}{Phys. Lett.} \textbf{\bibinfo{volume}{B110}},
  \bibinfo{pages}{35} (\bibinfo{year}{1982}).

\bibitem[{\citenamefont{Khoury et~al.}(2001)\citenamefont{Khoury, Ovrut,
  Steinhardt, and Turok}}]{Khoury:2001wf}
\bibinfo{author}{\bibfnamefont{J.}~\bibnamefont{Khoury}},
  \bibinfo{author}{\bibfnamefont{B.~A.} \bibnamefont{Ovrut}},
  \bibinfo{author}{\bibfnamefont{P.~J.} \bibnamefont{Steinhardt}},
  \bibnamefont{and} \bibinfo{author}{\bibfnamefont{N.}~\bibnamefont{Turok}},
  \bibinfo{journal}{Phys. Rev.} \textbf{\bibinfo{volume}{D64}},
  \bibinfo{pages}{123522} (\bibinfo{year}{2001}), \eprint{hep-th/0103239}.

\bibitem[{\citenamefont{Nielsen}(1978)}]{Nielsen:1978mp}
\bibinfo{author}{\bibfnamefont{N.~K.} \bibnamefont{Nielsen}},
  \bibinfo{journal}{Nucl. Phys.} \textbf{\bibinfo{volume}{B140}},
  \bibinfo{pages}{499} (\bibinfo{year}{1978}).

\bibitem[{\citenamefont{Kallosh}(1978)}]{Kallosh:1978de}
\bibinfo{author}{\bibfnamefont{R.~E.} \bibnamefont{Kallosh}},
  \bibinfo{journal}{Nucl. Phys.} \textbf{\bibinfo{volume}{B141}},
  \bibinfo{pages}{141} (\bibinfo{year}{1978}).

\bibitem[{\citenamefont{Freed and Moore}(2006)}]{Freed:2004yc}
\bibinfo{author}{\bibfnamefont{D.~S.} \bibnamefont{Freed}} \bibnamefont{and}
  \bibinfo{author}{\bibfnamefont{G.~W.} \bibnamefont{Moore}},
  \bibinfo{journal}{Commun. Math. Phys.} \textbf{\bibinfo{volume}{263}},
  \bibinfo{pages}{89} (\bibinfo{year}{2006}), \eprint{hep-th/0409135}.

\bibitem[{\citenamefont{Moss and Silva}(1997)}]{Moss:1996ip}
\bibinfo{author}{\bibfnamefont{I.~G.} \bibnamefont{Moss}} \bibnamefont{and}
  \bibinfo{author}{\bibfnamefont{P.~J.} \bibnamefont{Silva}},
  \bibinfo{journal}{Phys. Rev.} \textbf{\bibinfo{volume}{D55}},
  \bibinfo{pages}{1072} (\bibinfo{year}{1997}), \eprint{gr-qc/9610023}.

\bibitem[{\citenamefont{Lukic and Moore}(2007)}]{Lukic:2007aj}
\bibinfo{author}{\bibfnamefont{S.}~\bibnamefont{Lukic}} \bibnamefont{and}
  \bibinfo{author}{\bibfnamefont{G.~W.} \bibnamefont{Moore}}
  (\bibinfo{year}{2007}), \eprint{hep-th/0702160}.

\end{thebibliography}

\end{document}